\documentclass[fleqn,10pt]{wlscirep}
\usepackage[utf8]{inputenc}
\usepackage[T1]{fontenc}
\usepackage{amsmath}
\usepackage{bm}
\usepackage{amsthm}
\usepackage{amssymb}
\usepackage{mathrsfs}
\usepackage{xcolor}
\usepackage{chemformula}
\usepackage{multirow}
\usepackage{float}
\usepackage{graphicx} 
\usepackage{tabularx}
\usepackage{physics}
\usepackage{braket}
\usepackage[ruled, linesnumbered]{algorithm2e}
\usepackage[version=4]{mhchem}
\usepackage[caption=false]{subfig}
\usepackage{booktabs}
\usepackage[normalem]{ulem}

\newcommand{\nelec}{N_{\textrm{elec}}}

\newcommand{\rev}[1]{\textcolor{black}{#1}}

\title{\rev{A Hybrid Quantum Computing Pipeline for Real World Drug Discovery}}

\author[1$\dagger$]{Weitang Li}
\author[2, 4$^*\dagger$]{Zhi Yin}
\author[2$\dagger$]{Xiaoran Li}
\author[2$\dagger$]{Dongqiang Ma}
\author[2$\dagger$]{Shuang Yi}
\author[1]{Zhenxing Zhang}
\author[1]{Chenji Zou}
\author[1]{Kunliang Bu}
\author[1]{Maochun Dai}
\author[3]{Jie Yue}
\author[2]{Yuzong Chen}
\author[3$^*$]{Xiaojin Zhang}
\author[1$^*$]{Shengyu Zhang}
\affil[1]{Tencent Quantum Lab, Shenzhen, 518057, China}
\affil[2]{AceMapAI Biotechnology, Suzhou, 215000, China}
\affil[3]{AceMapAI Joint Lab, China Pharmaceutical University, Nanjing, 211198, China}
\affil[4]{School of Science, Ningbo University of Technology, Ningbo, 315211, China}

\vspace{10pt}

\affil[$\dagger$]{these authors contributed equally to this work}

\affil[*]{Corresponding authors. E-mail: yinzhi@acemapai.com; zxj@cpu.edu.cn; shengyzhang@tencent.com}

\begin{abstract}
Quantum computing, with its superior computational capabilities compared to classical approaches, holds the potential to revolutionize numerous scientific domains, including pharmaceuticals. However, the application of quantum computing for drug discovery has primarily been limited to proof-of-concept studies, which often fail to capture the intricacies of real-world drug development challenges. 
In this study, we diverge from conventional investigations by developing \rev{a hybrid} quantum computing pipeline tailored to address genuine drug design problems. \rev{Our approach underscores the application of quantum computation in drug discovery and propels it towards more scalable system.} We specifically construct our versatile quantum computing pipeline to address two critical tasks in drug discovery: the precise determination of Gibbs free energy profiles for prodrug activation involving covalent bond cleavage, and the accurate simulation of covalent bond interactions.
This work serves as a pioneering effort in benchmarking quantum computing against veritable scenarios encountered in drug design, especially the covalent bonding issue present in both of the case studies, thereby transitioning from theoretical models to tangible applications. Our results demonstrate the potential of a quantum computing pipeline for integration into real world drug design workflows. 
\end{abstract}
\begin{document}

\flushbottom
\maketitle
%
%
\thispagestyle{empty}

\section{Introduction}

Quantum computing is emerging as a powerful change that promises to significantly enhance scientific computing and simulations. Quantum computers, operating with quantum bits (qubits),  have the potential to execute complex calculations at speed and levels of precision that traditional supercomputers cannot achieve~\cite{arute2019quantum, zhong2020quantum, kim2023evidence}. The realm of drug discovery, characterized by its need for meticulous molecular modeling and predictive analytics~\cite{durrant2011molecular,sliwoski2014computational,ou2012computational,dara2022machine}, stands as an ideal candidate to benefit from this quantum leap. Recent endeavors  have commenced the integration of quantum computing into drug design research, marking a progressive stride in the application of advanced computational technologies to drug discovery \cite{wong2022fast,anton2021resource,batra2021quantum, outeiral2021prospects}. In drug design, existing classical computational chemistry methods are not able to compute exact solutions, and the required computational cost grows exponentially as the scale of the system grows. Quantum algorithms exemplified by the Variational Quantum Eigensolver (VQE)~\cite{tilly2022variational}, hold the potential to advance classical methods like Hartree-Fock (HF)~\cite{levine2009quantum} towards more accurate solutions within the quantum computing paradigm. As the scale of quantum computers expands, quantum computing approaches are expected to significantly outperform existing solutions, such as Density Functional Theory (DFT)~\cite{kohn1965self}, in terms of both accuracy and efficiency in scenarios involving quantum chemical calculations.
\rev{
In addition to the quantum chemistry approach, 
a variety of drug-design problems can be cast into optimization problems~\cite{dill2012protein, outeiral2021prospects}.
The quantum approximate optimization algorithm~\cite{robert2021resource} or quantum annealing algorithms~\cite{perdomo2012finding, marchand2019variable} can then be employed to solve these optimization algorithms.
}

\begin{figure*}[htbp]
    \includegraphics[width=1\textwidth]{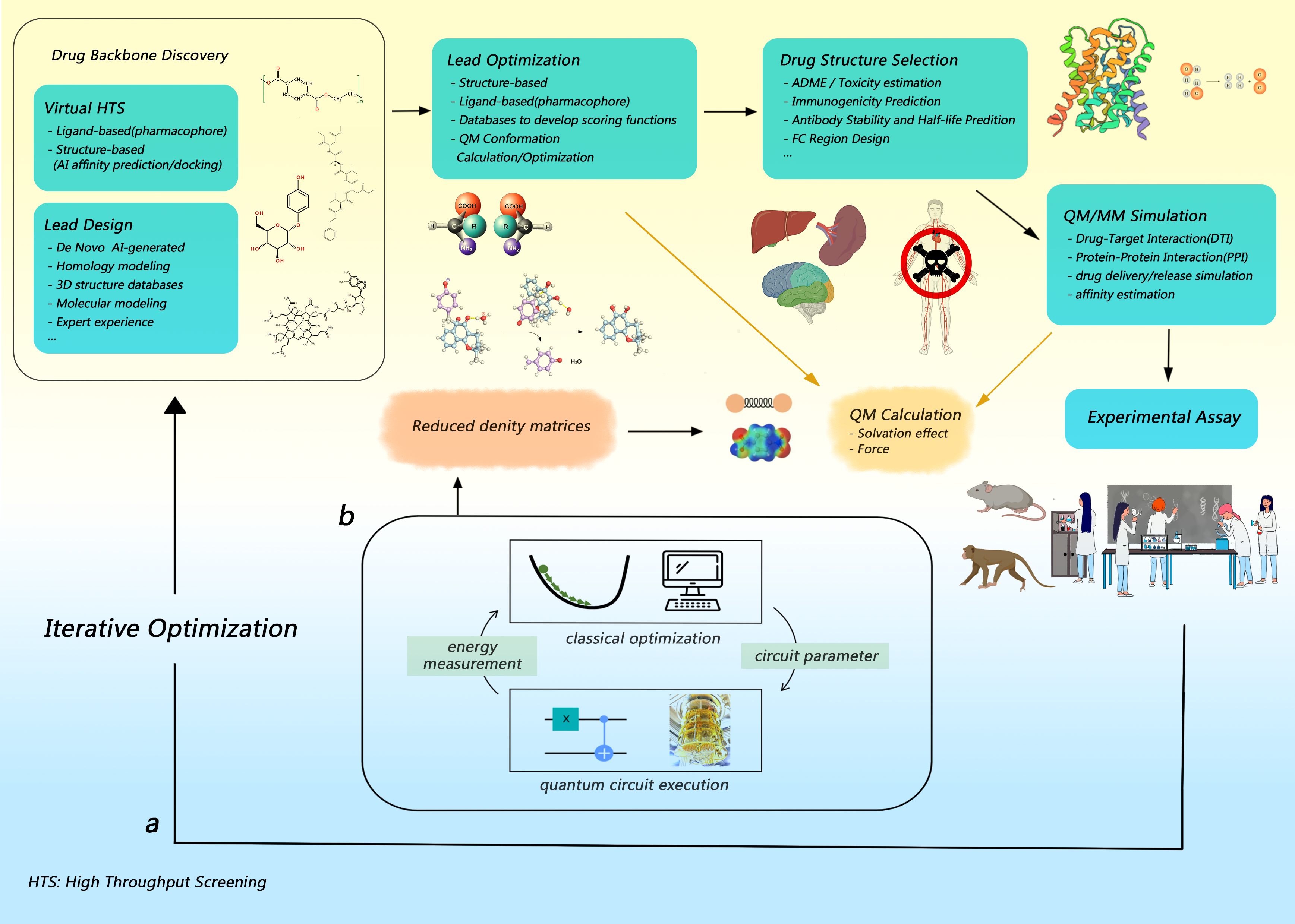}
    \caption{Schematic demonstration for the generalizable quantum computing pipeline for drug discovery. \textbf{a.} The standard workflow of computer-aided drug design (CADD). \textbf{b.} The module detailing the quantum computing process involved.}
    \label{fig:workflow}
\end{figure*}

However, in the current landscape, the involvement of quantum computing in drug discovery is primarily restricted to conceptual validation, \rev{with minimal integration into real world drug design\cite{santagati2024drug,otten2022localized,lau2023insights,gircha2023hybrid,blunt2022perspective,lam2020applications}.} \rev{Our hybrid quantum computing pipeline (see Figure \ref{fig:workflow}) is real-world drug discovery problem oriented. Our approach addresses this gap by investigating two pertinent case studies rooted in actual clinical and pre-clinical contexts.} 
The key step for quantum computation of molecular properties is to prepare the molecular wave function on a quantum device. To this end, the VQE framework is suitable for near-term quantum computers. 
As shown in Fig.~\ref{fig:workflow}(b), the core of VQE is to employ parameterized quantum circuits to measure the energy of the target molecular system. Then, a classical optimizer is employed to minimize the energy expectation until convergence. Due to the variational principle, the state of the quantum circuit becomes a good approximation for the wave function of the target molecule, and the measured energy is the variational ground state energy.
After that, additional measurements can be performed on the optimized quantum circuit for other interested physical properties.

Our first case study focuses on a carbon-carbon bond cleavage prodrug strategy\cite{gong_carboncarbon_2022}, which investigates an innovative prodrug activation approach applied to $\beta$-lapachone for cancer-specific targeting and has been validated through animal experiments. This prodrug design primarily aims to address the limitations of active drugs in pharmacokinetics and pharmacodynamics, offering a valuable supplement to the existing prodrug strategies\cite{zhou2020paclitaxel,rautio2018expanding,dong2020general,liu2022smart,luo2021activatable,weng2020harnessing,chang2021prodrug}. The simulation of the prodrug activation process requires precise modeling of the solvation effect in human body. To achieve this, we implement a general pipeline that enables quantum computing of the solvation energy based on the polarizable continuum model (PCM). Our findings demonstrate the viability of quantum computations in simulating covalent bond cleavage for prodrug activation calculations, which are important steps in real-world drug design tasks.

We then turn to the covalent inhibition of KRAS (Kirsten rat sarcoma viral oncogene), a protein target prevalent in numerous types of cancers. KRAS plays a crucial role in the RAS/MAPK (Mitogen-Activated Protein Kinase) signaling pathway, significantly influencing cell growth, differentiation, and survival \cite{huang_kras_2021}. Mutations of this protein, particularly the G12C variant, are common in various cancers, including lung and pancreatic cancers, and are associated with uncontrolled cell proliferation and cancer progression \cite{z_gtp-state-selective_2020, kim_pan-kras_2023, nikolaev_somatic_2018, pan_review_2021}. Sotorasib (development code name AMG 510), a covalent inhibitor targeting this mutation, has demonstrated potential in providing a more prolonged and specific interaction with the KRAS protein, a crucial approach in cancer therapy \cite{canon_clinical_2019,lanman_discovery_2020}. Since the introduction of AMG 510, a flurry of new inhibitors targeting G12C has been developed, and even expanding to other KRAS mutations \cite{nagasaka_kras_2020, fell_identification_2020,wang_identification_2022,nakayama_characterisation_2022,tanaka_clinical_2021}, and several candidates for broad spectrum inhibition have also been proposed \cite{cheng_novel_2023}. However, the other mutations usually don't have a potential site for covalent binding, so their efficacy remained to be rigorously tested.

Quantum computing can enhance our understanding of such drug-target interactions through QM/MM (Quantum Mechanics/Molecular Mechanics) simulations, which are vital in the post-drug-design computational validation phase. To realize this, \rev{we implemented a hybrid quantum computing workflow for molecular forces during QM/MM simulation.} This development not only facilitates a detailed examination of covalent inhibitors like Sotorasib, but also propels the field of computational drug development forward.

Through these two real-world drug design examples, \rev{we present a hybrid quantum computing pipeline for drug design.} Our workflow has advantages in its flexibility and has been carefully constructed to accommodate various applications in the area of drug discovery. The universality of our pipeline highlights its potential as a foundational tool, empowering researchers with a ready-to-use computational resource. 

En route of our computational investigations, we also established a number of benchmarks, which not only exemplify the robustness of our approach but also serve as a valuable reference for the field of quantum computing-enhanced drug discovery. By democratizing access to this advanced pipeline, we lay the groundwork for expanded collaborative endeavors within the scientific community, thereby accelerating the translation of quantum computing power into tangible therapeutic outcomes.

\section{Results}

\subsection{Gibbs Free Energy Profiles in Prodrug Activation Strategy}

\subsubsection{Carbon-carbon bond cleavage in prodrug activation strategy}

In modern drug research, prodrug activation is a very important strategy\cite{ferrara2020cns,xia2021prodrug}. It helps turn inactive ingredients into active drugs inside the body. This strategy helps make drugs work better by making sure they only activate at certain places in the body, which lowers the risk of side effects and leads to safer and more effective treatments.

Among various prodrug activation strategies\cite{gong_carboncarbon_2022,zhou2020paclitaxel,rautio2018expanding,dong2020general,liu2022smart,luo2021activatable,weng2020harnessing,chang2021prodrug}, that based on the cleavage of carbon-carbon (C-C) bonds is particularly innovative. It is a novel strategy with applicability to drugs without traditional modifiable groups. The C-C bond, a quintessential linkage in organic chemistry, imparts robustness to molecular frameworks, and its selective scission demands conditions of exquisite precision. Synthesizing prodrugs that are primed for C-C bond cleavage under pathophysiological conditions confronts us with the dual challenges of sophisticated synthetic chemistry and intricate mechanistic elucidation.

\begin{figure*}[htbp]
    \centering
    \includegraphics[width=1\textwidth]{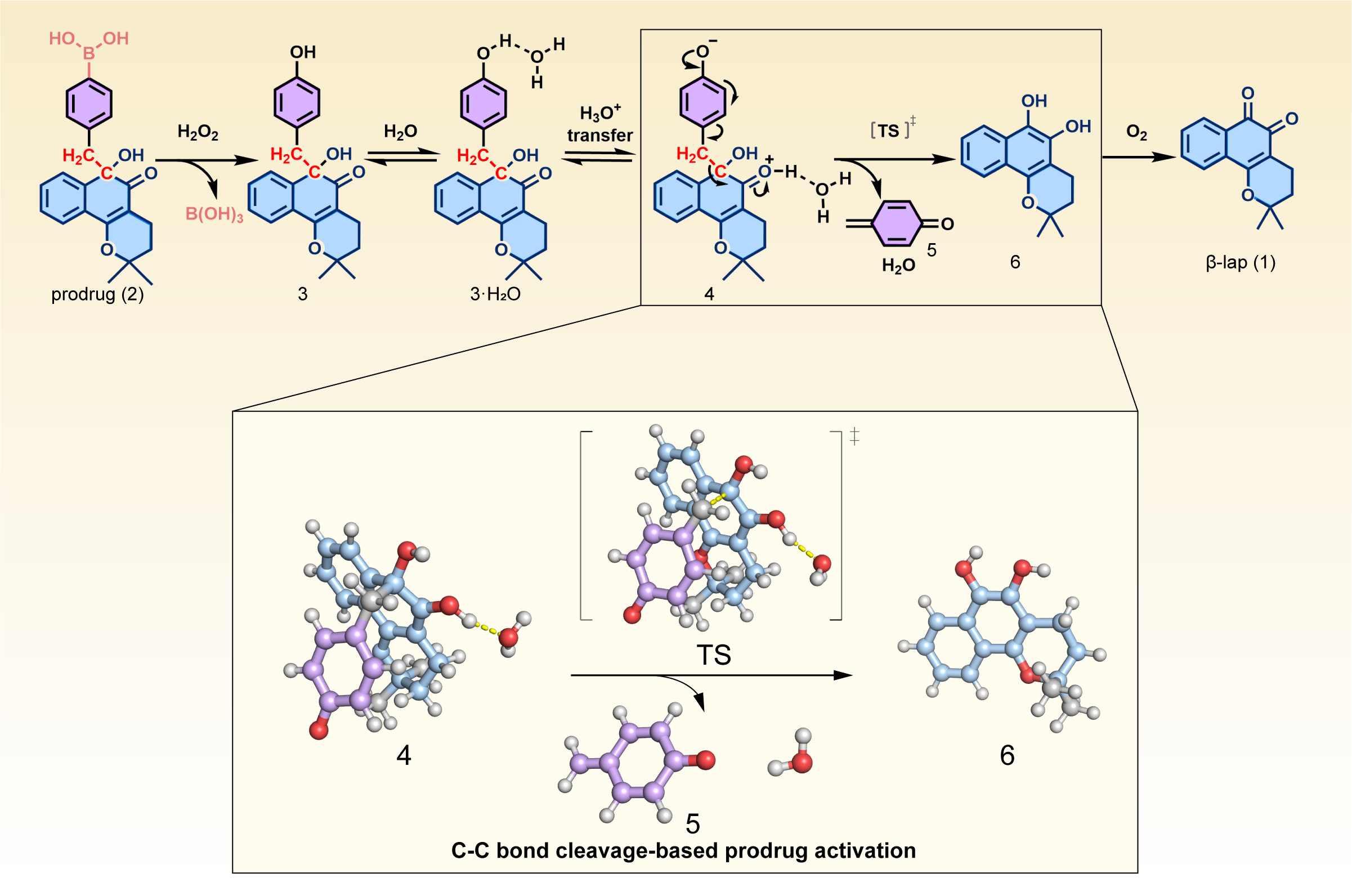}
    \caption{Schematic illustration of components(\textbf{4}, \textbf{5}, \textbf{6} and \textbf{TS}) involved in the process of the C-C bond cleavage-based activated drug release. For ease of comparison, we have adopted the molecular numbering from the original work of carbon-carbon bond cleavage based prodrug activation strategy.}
    \label{fig:ccbond}
\end{figure*}

In this cleavage of carbon-carbon bonds based prodrug activation strategy, the calculation of the energy barrier is crucial because it determines whether the chemical reaction can proceed spontaneously under physiological conditions. It also plays a significant role in determining stable molecular structures, guiding molecular design, and evaluating molecular dynamic properties. To simplify the computations, in the subsystem where quantum computing is employed, we have selected five key molecules involved in the cleavage of the C-C bond as simulation subjects, 
performing the single-point energy calculation and the essential solvent model calculations after conformational optimization process (see Figure \ref{fig:ccbond} for details). Considering the practical value and significance of prodrug activation strategies in current drug design, especially for drug delivery, our calculations are suitable for extension to more similar scenarios.

\subsubsection{Gibbs free energy profiles of covalent bond cleavage for prodrug activation}

Gibbs free energy profiling of covalent bond cleavage is a critical task for drug design, especially prodrug activation. It is of great importance for the selectivity and efficacy of therapeutic agents, guiding synthetic routes, and even achieving accurate molecular models for complex chemical reactions.

In this work, we study the prodrug design for $\beta$-lapachone, a natural product with extensive anticancer activity.
In the original study~\cite{gong_carboncarbon_2022}, the authors use DFT and select M06-2X functional to calculate the energy barrier. The results show that the energy barrier for C-C bond cleavage is small enough for the chemical reaction to proceed spontaneously under physiological temperature conditions. It's worth mentioning that in the original study, this novel prodrug design strategy is validated through wet laboratory experiments. In this study, we employed two classical computational methods, namely HF and Complete Active Space Configuration Interaction (CASCI), to compute reference values for quantum computation. While DFT is typically the preferred method in conventional pharmacochemical reaction calculations due to its efficiency and accuracy, the choice of HF and CASCI methods in this study yields reaction barrier that is consistent with wet lab results.

\begin{figure}[H]
    \centering
    \includegraphics[width=0.6\linewidth]{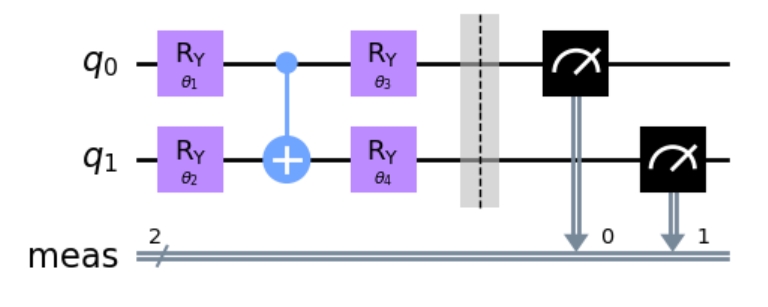}
    \caption{The 2-qubit quantum circuit used in this study. The state of the quantum circuit is adjusted by 4 parameterized $R_y$ gates.}
    \label{fig:circuit}
\end{figure}

Despite that quantum devices with more than 100 qubits are becoming available, simulating large chemical systems would require very deep circuits, which will inevitably lead to inaccurate outcomes due to intrinsic quantum noise.
Additionally, the $N^4$ terms to measure to calculate molecular energy is another bottleneck for quantum computation due to the limited measurement shot budget.
Thus, it is often desirable to reduce the effective problem size of chemical systems, so that they can be processed on available quantum devices.
The quantum embedding methods and downfolding methods have gathered significant attention recently~\cite{li2022toward, Kowalski2023}.
In this work, we employed the active space approximation due to its popularity and versatility,
which simplifies the QM region into a more manageable 2 electron/2 orbital system.
\rev{The CASCI energy can be considered as the exact solution under the active space approximation and the results by quantum computers are expected to be consistent with the CASCI energy.}
The fermionic Hamiltonian is then converted into a qubit Hamiltonian using parity transformation. 
The wave function of the active space can then be represented by a 2-qubit superconducting quantum device. We utilized a hardware-efficient $R_y$ ansatz with a single layer as the parameterized quantum circuit for VQE, as depicted in Fig.~\ref{fig:circuit}. 
We applied standard readout error mitigation to enhance the accuracy of the measurement results.
For more detailed technical information, please refer to the Methods section. We implemented the entire workflow in the \textsc{TenCirChem} package~\cite{li2023tencirchem}, allowing users to utilize these functions with just a few lines of code.

By calculating the energy barrier for C-C bond cleavage, we compare our quantum computing results with those from the original study. Our computation involves single-point energy calculations with the influence of water solvation effects.
For both classical and quantum computations, we selected the 6-311G(d,p) basis set and chose the ddCOSMO model as the solvation model. The thermal Gibbs corrections were calculated at the HF level.
Additionally, we included the results from HF and CASCI, which are based on classical computational chemistry, for comparison.
In  Table~\ref{tab:delta_energy} we list the reaction barrier $\Delta G^{\ddag}$ and the reaction Gibbs free energy change $\Delta G$ for the prodrug-activation reaction.
The Gibbs free energy for relevant molecules is listed in Table~\ref{tab:energy_data} in the Methods section.
The results of the reaction energy barrier $\Delta G^{\ddag}$ obtained from both classical quantum chemistry calculation methods and quantum computing methods are in good agreement.
They also align closely with the calculation results of the original paper, which employed the M06-2X functional and Gaussian as the computational tool.
From the activation barrier results from quantum computers in Table \ref{tab:delta_energy}, we observe that the activation barrier is less than 20 kcal/mol. In the field of drug design, this indicates that the reaction could spontaneously occur within a biological organism. Therefore, the results from quantum computers in our pipeline can be used for the assessment of prodrug activation processes.
On the other hand, we obtained significantly lower energy values $\Delta G$ compared to the DFT method in the original study.
It is worth noting that without considering the solvation effect, both HF and CASCI calculations yield much lower reaction barriers $\Delta G^{\ddag}$.
In fact, the VQE method even produces a non-physical negative reaction barrier.
This observation emphasizes the importance of considering the solvation effect in the drug-design pipeline.

The similarity between the results obtained from HF, CASCI, and VQE can be attributed to the relatively small active space considered in this study. 
There are studies indicating that quantum computational methods like VQE can achieve near-exact solutions for medium-sized chemical systems~\cite{lee2018generalized, grimsley2019adaptive}. As the scale of quantum computing continues to grow, we may be able to alleviate the active space approximation employed in this work and make significant improvements to the HF method. Our results demonstrate the effectiveness of quantum computing in scenarios involving Gibbs free energy profile calculations of covalent bond cleavage, as well as the versatility and plug-and-play advantages of our pipeline.

Next, we discuss the computational wall time required for quantum computation. 
In the (2e, 2o) active space, the bottleneck for both classical and quantum computation
is obtaining the HF solution with the solvation effect.
Thus, the total wall times are comparable for all molecules computed in this study,
ranging from several minutes to approximately one hour, depending on the size of the molecule.
The time cost for solving (2e, 2o) active space does vary between CPU and QPU,
as illustrated in Table \ref{tab:time_cmp}. 
Taking molecule $\textbf{5}$ as an example, classical computers require 3 seconds to complete the computation.
On the other hand, quantum computers take 63 seconds to perform the computation, and the majority of the time is dedicated to measuring the active space energy and the one-body reduced density matrix for the solvation effect.
Since active resetting is not implemented yet, for each measurement shot, the quantum computational bottleneck is to wait several times the decay ($T1$) time so that the energy stored in qubits is relaxed into the environment.
This results in an approximate duration of 1 ms for each measurement shot.
To determine the expectation value for each Pauli operator, 8192 measurement shots are performed, corresponding to a duration of 8 seconds.
For energy evaluation, there are 8 Pauli strings to be measured, which are grouped into 5 measurement groups based on commutation relations.
As a result, calculating the energy expectation takes approximately 40 seconds.
The calculation of one-body reduced density matrices in the active space involves measuring 3 additional expectation values.
Thus, the total time cost for the quantum computing kernel is approximately 60 seconds,  consistent with our experimental findings.
Although the active space size remains the same for different molecules, the time cost for classical and quantum computation does vary.
For example, calculating \textbf{4} and \textbf{TS} is significantly more time-costly than computing \textbf{5}.
This discrepancy arises from the differing time required for active space integral transformation across molecules.
Nevertheless, for all molecules, quantum computation takes approximately one minute longer than classical computers.

In this study, we have limited the utilization of quantum computers to a few qubits employing
the active space approximation, due to the limited size and gate noise of currently available quantum computers.
Herein, we estimate what kind of quantum computers are required for a fully correlated computation of the systems studied in this work without incurring the active space approximation.
Taking molecule \textbf{4} as an example, with 6-311G(d,p) basis set, the system corresponds to $N=630$ orbitals and $\nelec=196$ electrons.
To reduce the qubit requirement, the paired unitary coupled-cluster ansatz can be employed, which requires only 1 qubits for each orbital due to the restriction of electron pairing~\cite{zhao2023orbital, o2023purification}.
Other advantages of the ansatz include that evaluating the energy requires only constant measurement
and linear circuit depth due to the efficient Givens-SWAP network.
Additionally, since there are $\nelec=196$ electrons, the number of all possible double excitations is $\frac{\nelec}{2}\times(N-\frac{\nelec}{2})=52136$.
Thus, a fully correlated computation at PUCC level with double excitations (PUCCD) involves a quantum circuit with approximately $10^3$ qubits and $10^5$ Givens-SWAP gates.
\rev{
The PUCCD ansatz has been successfully implemented on both superconducting and trapped-ion quantum computers~\cite{zhao2023orbital, o2023purification}.
The number of qubits employed in these studies is around 10, sufficient to describe 1 to 2 heavy atoms if active space or embedding techniques are not used~\cite{liu2022quantum}.
While digital quantum computers with over 100 qubits are accessible~\cite{kim2023evidence}, their application in quantum chemistry has been limited, primarily due to the restricted fidelity of two-qubit gates.
However, with improved two-qubit gate fidelity, these quantum computers can handle complex molecules comprising dozens of atoms, such as molecule \textbf{4}.
}

Empirically, there are $0.7\times N^2$ Pauli strings in the PUCC Hamiltonian, which leads to approximately $M=3\times10^5$ terms when $N=630$.
These terms can be divided into 3 measurement groups. 
Assuming for each group $K$ repeated circuits are executed for measurement,
the expectation variance $\epsilon^2$ is approximately $\frac{1}{K}\sum_j^M |\alpha_j|^2$.
Thus, if we wish to achieve the measurement precision to $\epsilon=0.01$ Hartree and $|\alpha_j|$ is assumed to be 0.1 Hartree,
the number of measurement shots $K$ is $10^7$ and the total number of shots for 3 measurement groups is approximately $10^8$.
On superconducting quantum computing platforms, the reset time is the bottleneck for circuit execution, which can be estimated as $10^{-3}$ seconds.
Thus, it takes $10^5$ seconds to measure the molecular energy, the key step for VQE.
Since computing the solvation energy requires only one-body reduced density matrix, the additional measurement cost can be neglected.
The multiplicative factor for parameter optimization is not considered.
If a set of accurate circuit initial parameters can be computed through classical preprocessing, such as quantum chemistry computation or machine learning~\cite{Romero19, jain2022graph}, we may conclude that using a single quantum processor it takes $10^5$ seconds to compute the solvation energy.
The $3K$ repeated circuits can be easily paralleled. In the optimal situation where $3K$ quantum processors are available for usage, the time cost for QM calculation can be reduced to $10^{-3}$ second.

\subsection{Covalent Bond Simulation}

\begin{figure}[H]
    \centering
    \subfloat[\label{fig:6oim_sterio}]{
        \includegraphics[width=0.48\textwidth]{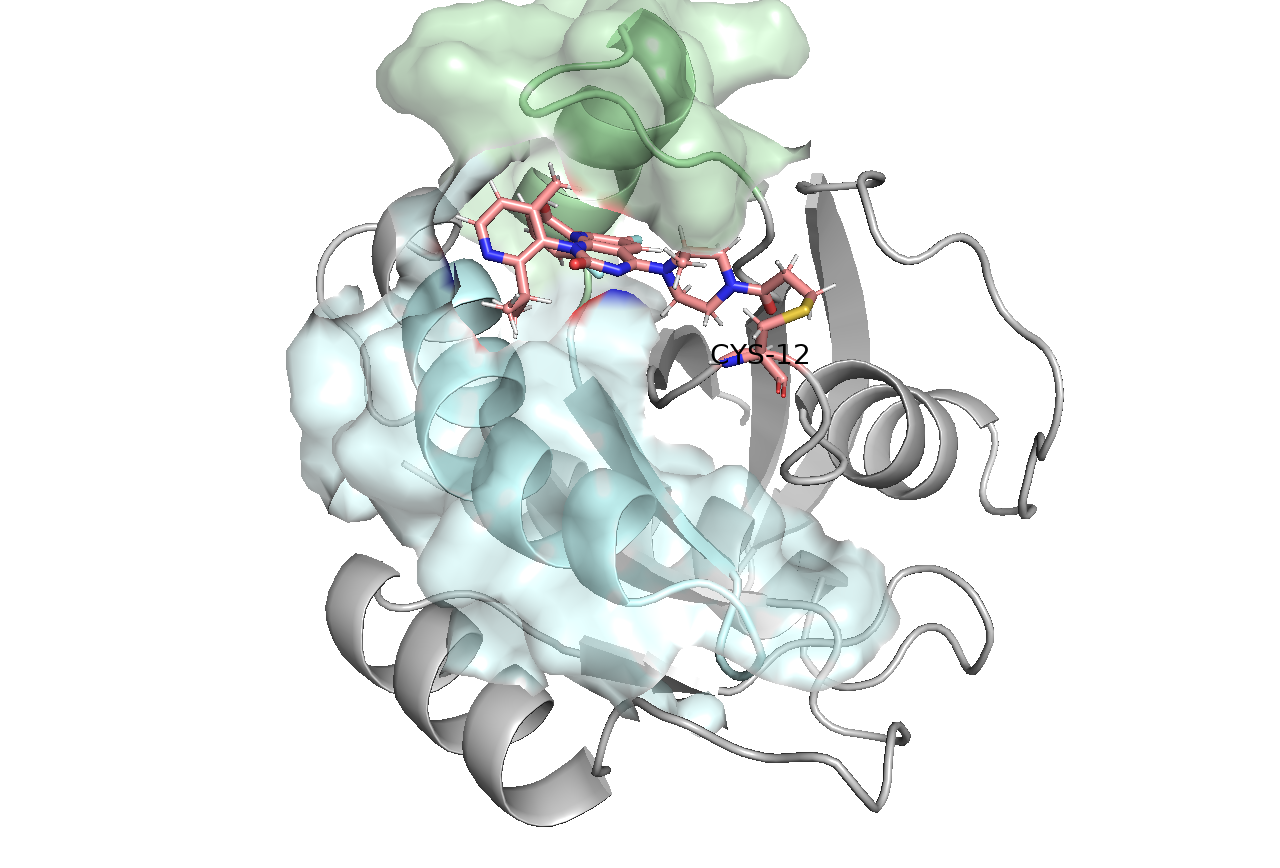}
    }
    \subfloat[\label{fig:6oim_atoms}]{
        \includegraphics[width=0.48\textwidth]{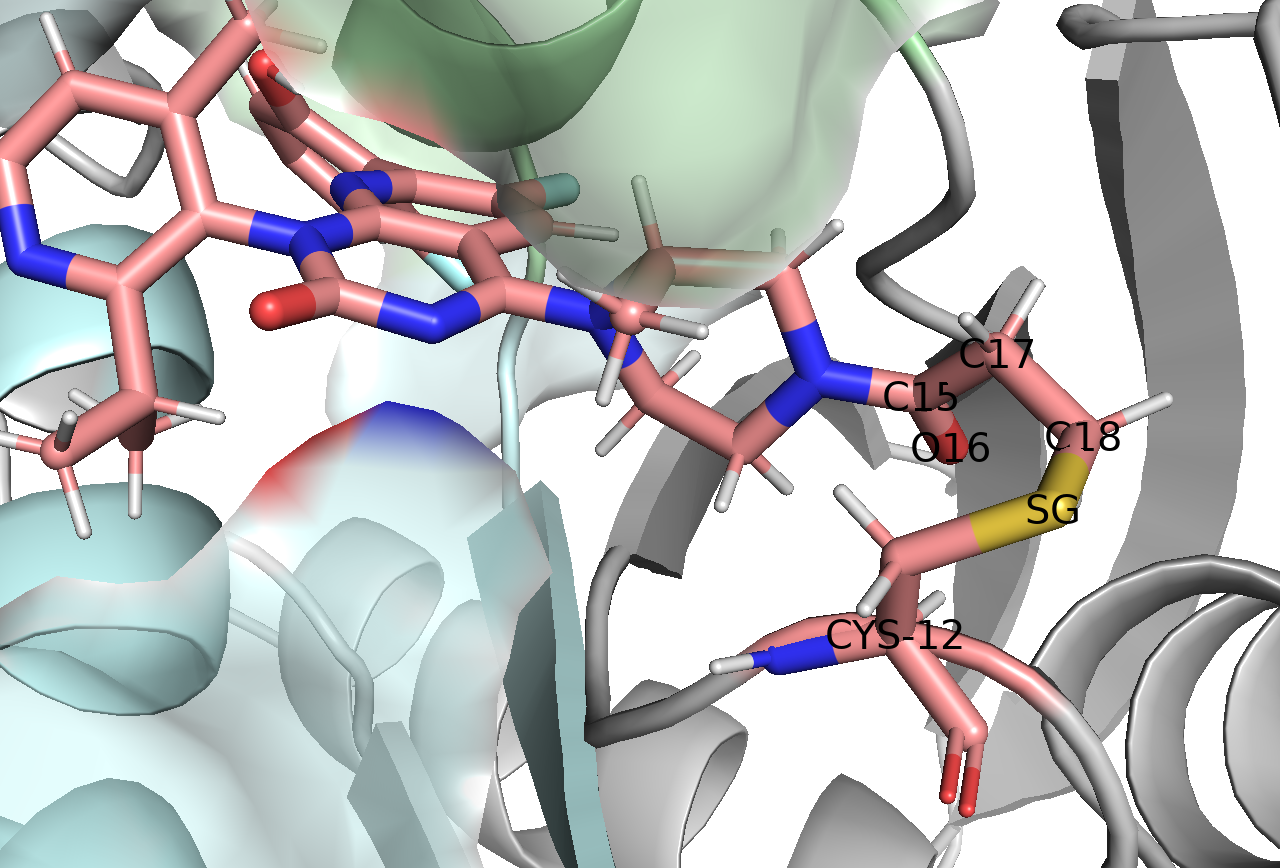}
    }
    \caption{Left: The KRAS-Sotorasib bonded structure. The cystein-Sotorasib part is shown as sticks while the rest of the system as ribbons. Right: Choosing the QM region. The atoms labeled SG, C18, C17, O16, and C15 are chosen.}
    \label{fig:6oim}
\end{figure}

KRAS is a prominent target in cancer therapy due to its significant role in various cancers, and the G12C mutation has been its most frequent and consequential mutation. The Sotorasib, an innovative covalent inhibitor targeting this mutation, represents a paradigm shift to KRAS-related cancer treatment. We set up a QM/MM simulation framework for the target-inhibitor interaction, and chose the QM region carefully to cover the key atoms involved in the covalent bond formation (see Figure \ref{fig:6oim} for a schematic exposition). 
We first run the QM/MM simulation on classical computers to get the baseline statistics, then move the QM energy computation to quantum computers and make sure that we can get comparable results.
The same with the case study for prodrug activation, a (2e, 2o) active space approximation is employed to reduce the measurement cost, and the active space wavefunction is processed using 2 qubits.

\subsubsection{KRAS and covalent inhibition}

\begin{figure}[H]
    \centering
    \subfloat[\label{fig:6oim_md}]{
        \includegraphics[width=0.48\textwidth]{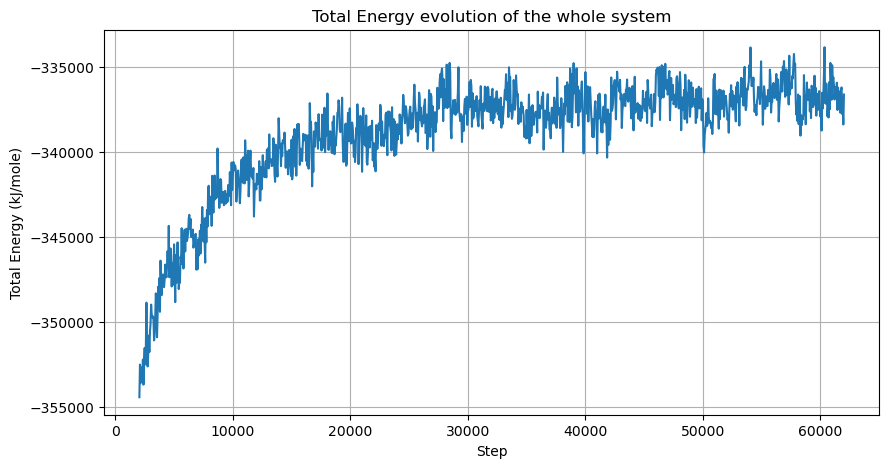}
    }
    \subfloat[\label{fig:6oim_qm}]{
        \includegraphics[width=0.48\textwidth]{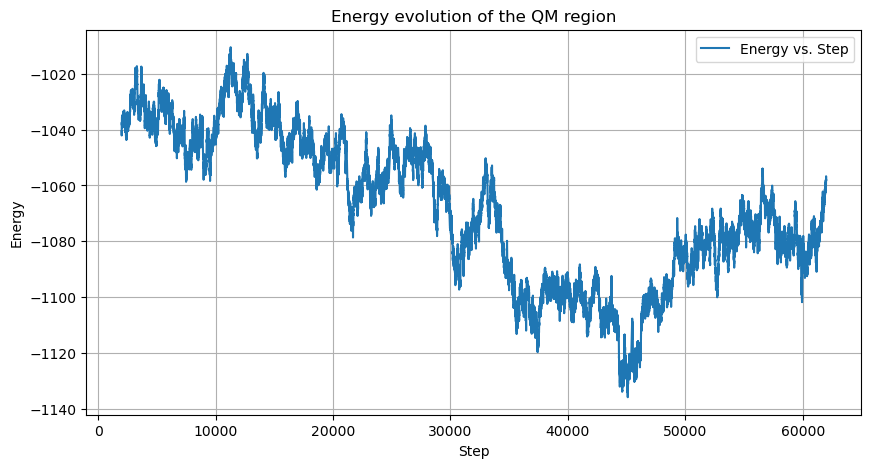}
    }
    
    \medskip
    
    \subfloat[\label{fig:6oim_bond}]{
        \includegraphics[width=0.48\textwidth]{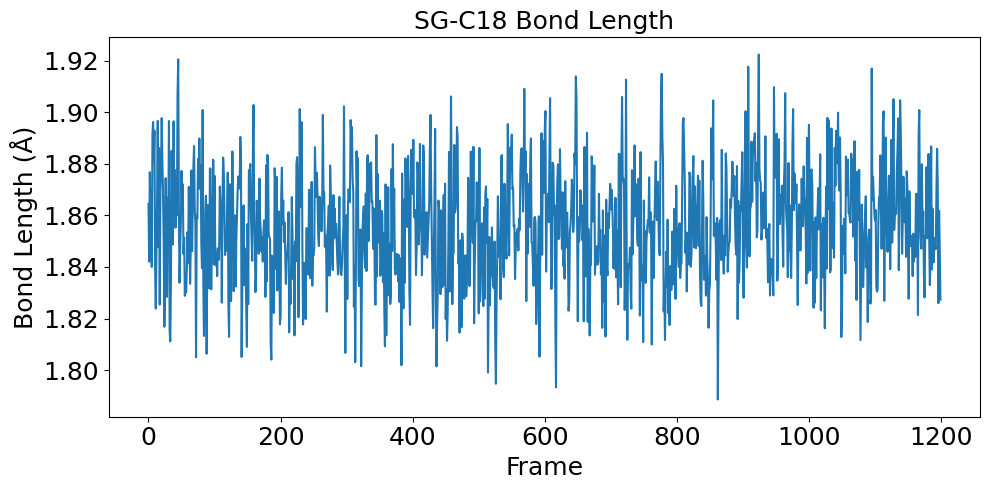}
    }
    \subfloat[\label{fig:6oim_angle}]{
        \includegraphics[width=0.48\textwidth]{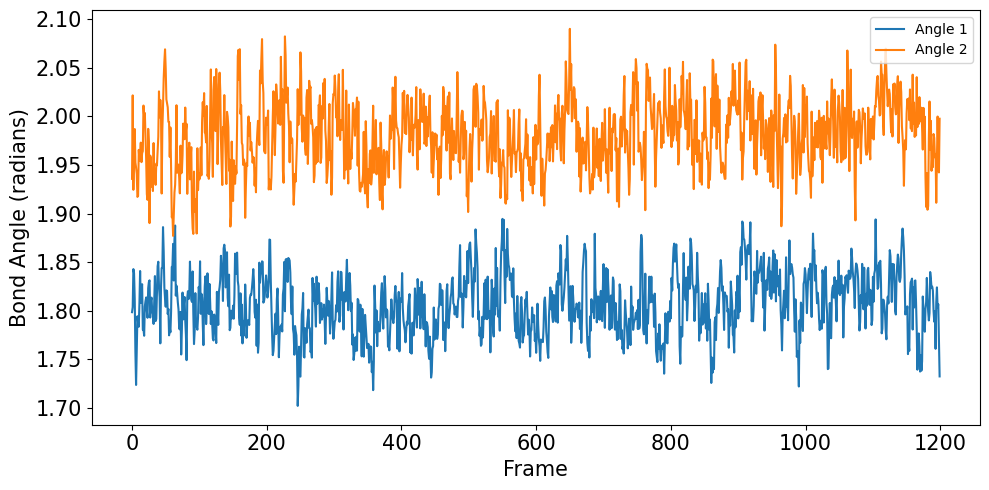}
    }
    \caption{(a): Energy evolution during the MD simulation. The energy stabilized after an initial equilibrating phase. (b): Monitoring the QM region energy evolution. (c) The covalent bond is remarkably stable during the whole simulation process. The bond length fluctuates around 1.86 angstrom with a standard deviation less than 0.1 angstrom. The bond length is aligned with previous literature discoveries, and the small deviation is a good indication of the bonding stability.(d) \rev{Visualising the bond angle variations during the simulation (CB-SG-C18 and SG-C18-C17).}}
    \label{fig:6oim_energy}
\end{figure}

To establish a robust baseline for the later quantum computer adaptation, close supervision of the
energy evolution of the QM region was conducted throughout the simulation,
as shown in Figure \ref{fig:6oim_qm}. Complementarily, the MM region encompassed the larger protein environment, including water
molecules and other cellular components, offering a realistic context for the
interaction. The energy transitions, including the potential energy, the kinetic
energy, and the system total energy, had been recorded, as shown in Figure \ref{fig:6oim_md}.

A critical reason that inhibiting KRAS had been so difficult, and the inhibition
of the KRAS G12C mutation had been so significant, is the possibility of designing small molecular inhibitors that specifically target the G12C mutation by
forming a covalent bond between the target and the inhibitor. For this reason,
it's imperative that Sotorasib can form a stable bonded complex with the target,
through covalent bonding. The bond length, bond angles, and dihedral angles
around the covalent bond had also been closely monitored during the simulation,
as shown in Fig \ref{fig:6oim_bond} and Figure \ref{fig:6oim_angle}.

We observed a specific and strong bond between Sotorasib and the target mutation,
offering critical insights into the drug's potential efficacy. This
understanding is pivotal for the rational design of future inhibitors targeting
similar mutations.

\subsubsection{QM energy update using quantum computers}

After establishing the QM/MM baseline, we then moved the QM computation first to a quantum emulator using TenCirChem, and then to a quantum computer.
The kernel of our calculation is again the VQE algorithm.
The MM region, represented as point charges, contributes a background potential to the Hamiltonian. 
The calculation of molecular forces is a common routine in classical computational chemistry.
Recent attempts have been made to transfer the algorithm to quantum computing platforms~\cite{o2019calculating, delgado2021variational, o2022efficient, sugisaki2022quantum, lai2023accurate}.
In our work,  the calculation is more complicated compared with previous studies, due to the active space approximation employed.
In addition, our work is the first example of integrating quantum computed forces into a full-scale QM/MM simulation workflow.
The details of the procedure are shown in the Methods section.

\begin{figure}[H]
    \centering
    \includegraphics[width=0.8\linewidth]{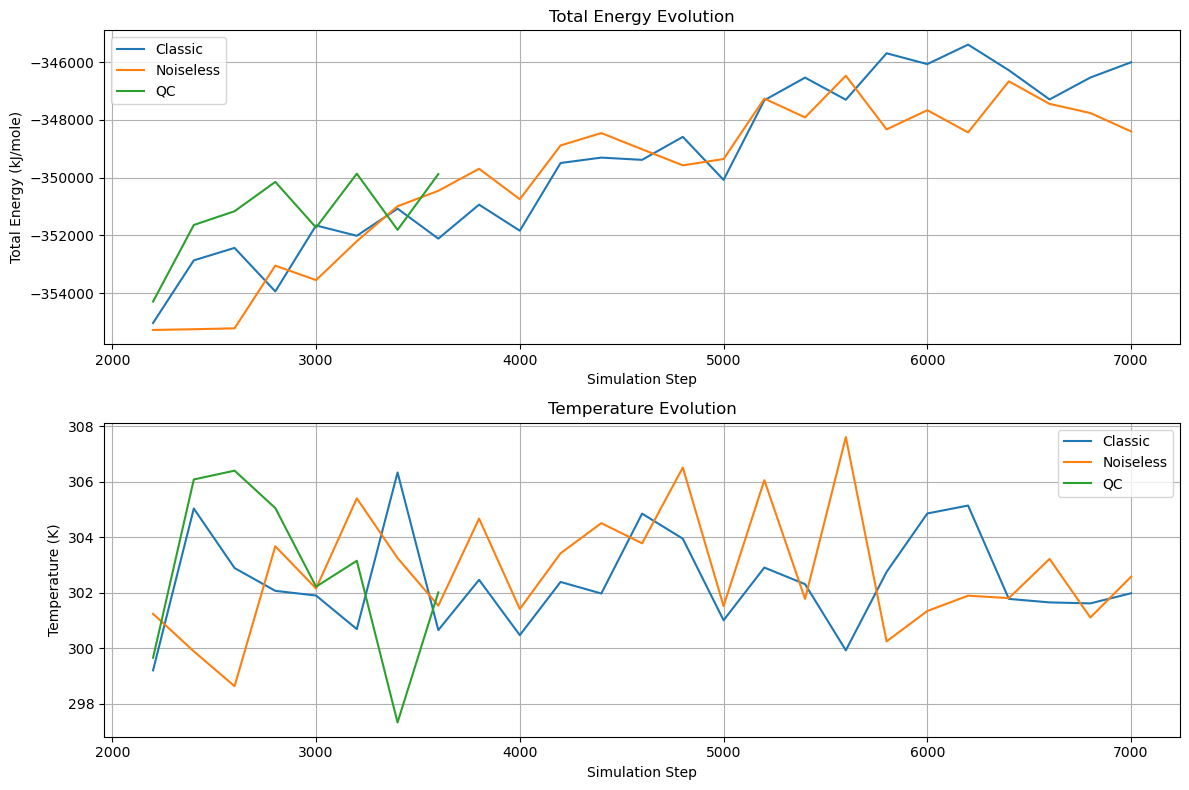}
    \caption{Energy transition of the classical simulation(in blue), the noiseless quantum emulation (in orange), and the quantum computer simulation(in green). The fluctuation falls neatly in the permissible deviations of the molecular dynamics simulation.}
    \label{fig:6oim_qc}
\end{figure}

\begin{figure}[H]
    \centering
    \subfloat[\label{fig:6oim_1228}]{
        \includegraphics[width=0.48\textwidth]{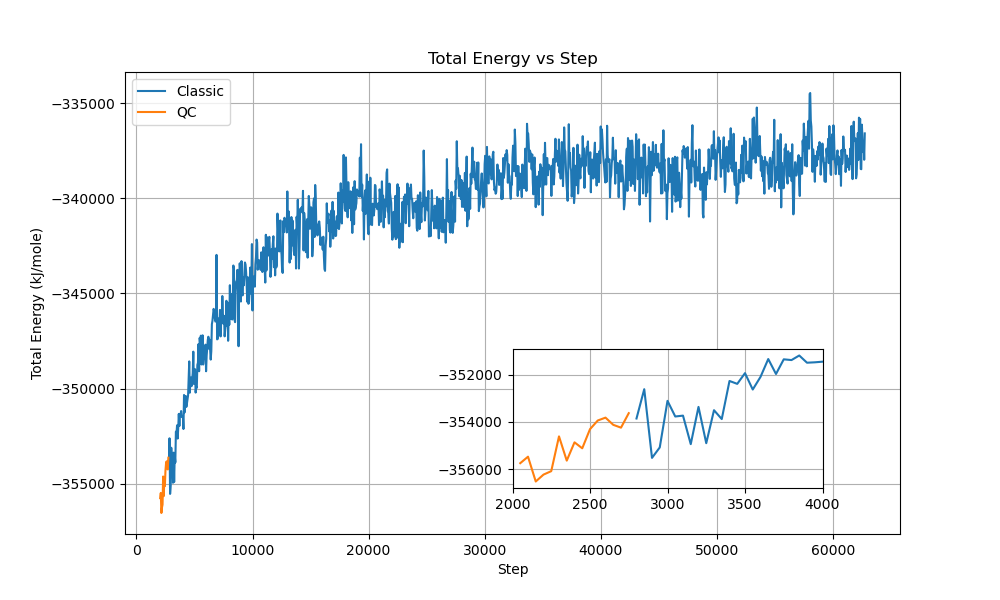}
    }
    \subfloat[\label{fig:6oim_1221}]{
        \includegraphics[width=0.48\textwidth]{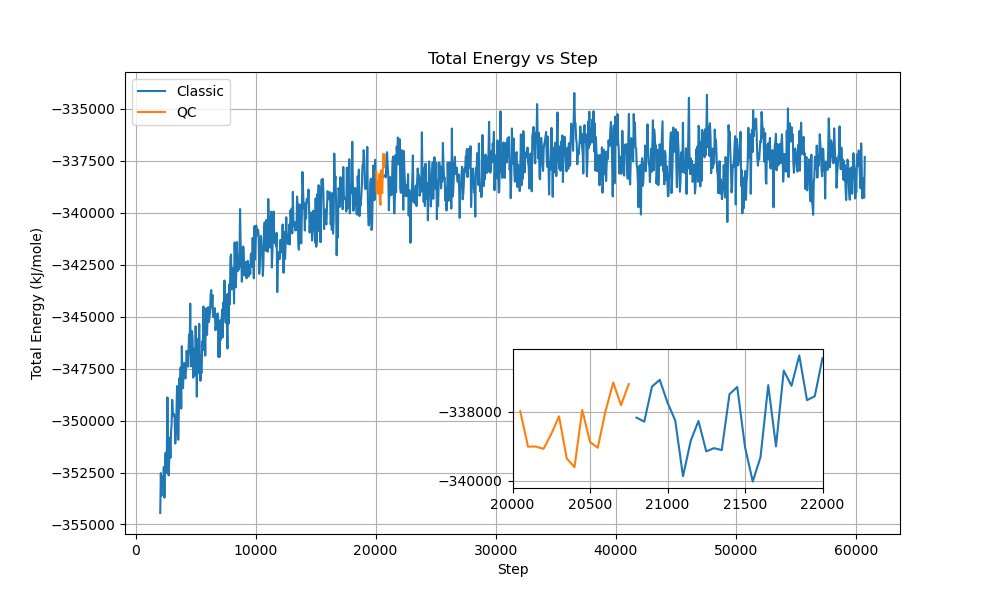}
    }
    \caption{Moving key computation to the Quantum computer. Left: the simulation is started on a quantum computer, and then moved to a classical machine. Right: the simulation is started on a classical machine, moved to a quantum computer halfway, and then moved back to a classical computer.}
    \label{fig:6oim_mix}
\end{figure}

Considering the computational load, to check the soundness of the computation, we run the first 1600 steps of the simulation on a quantum computer as a sanity check, and the results closely follow the baseline QM/MM simulation, as can be observed in Fig \ref{fig:6oim_qc}. We then moved some key steps of the QM/MM simulation to the quantum computer, to establish a QM/MM-QC hybrid simulation system. In Fig \ref{fig:6oim_1228}, the simulation is started on the quantum computer and continued on a classic computer; in Fig \ref{fig:6oim_1221}, the simulation is started on a classical computer, continued on a quantum machine, and subsequently moved back to the classic computer. Compared with the previous QM/MM simulation, we can see that the hybrid simulations have been able to closely follow the baseline trajectory, which gives us confidence that such hybrid simulations are a feasible use of the limited quantum computer computation powers.

The computational time cost comparison can be seen in Table \ref{tab:compare}. For classic QM/MM simulation, we utilized a high-performance system with dual Intel(R) Xeon(R) Gold 5220 CPUs (72 cores, 144 threads total, 2.20GHz base frequency), augmented by six NVIDIA A100-PCIe GPUs with 40,960 MiB memory each. The system is supported by 385 GB of RAM, facilitating the handling of extensive computational workloads.
Similar to the case study of prodrug activation, the time cost for quantum computers is larger than that for classical computers. 
To compute molecular forces, the two-body reduced density matrices need to be measured,
so at each step, the time cost is approximately two times the time cost for single-point energy calculation in Table~\ref{tab:time_cmp}.


The insights gained from these QM/MM simulations are not just
confined to the molecular interaction between Sotorasib and the KRAS(G12C)
protein. They lay the groundwork for future computations on a quantum computer,
promising to enhance the accuracy and speed of our drug discovery processes.
This step towards quantum computing implementation represents a transformative
progression in our research methodology, aligning with our ongoing efforts to
integrate advanced computational techniques in drug discovery.

Similar to the prodrug activation case, here for the covalent bond simulation case,
we provide an estimation of the quantum resource required for a fully correlated treatment of the QM region
using the pUCCD circuit.
The QM region is composed of 5 heavy atoms,
which are translated to $N=49$ orbitals with 6-31G basis set.
Thus a correlated computation without active space approximation requires a quantum circuit with approximately $50$ qubits and $588$ Givens-SWAP gates.
The total number of measurement shots is $10^6$ and $10^3$ seconds are required for an energy evaluation.
Since all elements of one and two-body reduced density matrices are also available from the three groups of measurement,
the wall time cost for the additional computation of molecular forces can be neglected.

\section{Discussions}

In this study, \rev{we have established a model pipeline that enables quantum computers to tackle real-world drug discovery problems.} 
Specifically, we have addressed two crucial challenges of computer-aided drug design, computing reaction barriers and molecular dynamics simulation.
Our pipeline combines quantum-classical hybrid computing platforms, leveraging the VQE framework on the quantum computing side to efficiently store and manipulate molecular wave functions. On the classical computing side, we employ the ddCOSMO solvation model to compute solvation energy and analytical CASCI force formula to compute molecular forces for QM/MM simulation, respectively. The interface between the quantum and classical computing sides relies on the one and two-body reduced density matrices.

To demonstrate the potential of our pipeline, we conducted two case studies using a superconducting quantum device. In the first case, we studied the Gibbs free energy profile for prodrug activation involving carbon-carbon bond cleavage under solvent conditions. The obtained reaction barrier and Gibbs energy change align well with previous experimental and theoretical studies. In the second case, we investigated a covalent inhibitor for KRAS(G12C) using QM/MM simulation. We closely monitored the evolution of energy and compared the time cost based on classical computers and quantum computers.

Based on the two cases, \rev{we provide evidence that our hybrid quantum computing pipeline has the potential to solve real-world drug design problems.} However, it is important to note that the accuracy of VQE calculations and the resources consumed require further improvement. On the quantum hardware side, continuous efforts should be made to enhance gate fidelity and strive toward achieving error correction. 
In terms of quantum algorithms, advanced VQE additions such as neural networks or Clifford circuits~\cite{zhang2022variational, shang2023schrodinger} can be explored to enhance the accuracy of the VQE circuit. 
While our study employed classical pre-optimization instead of parameter optimization on quantum computers due to associated overhead, the next step in the development of the VQE pipeline should involve better circuit parameter initialization and more efficient parameter optimization algorithms. This will enable the complete transfer of the pipeline onto quantum computers, further leveraging their computational power.

\rev{While there are plenty of works in leveraging quantum computing for drug discovery\cite{santagati2024drug,otten2022localized,lau2023insights,gircha2023hybrid,blunt2022perspective,lam2020applications}, the focus of our pipeline is for tackling specific real-world drug design problems. We emphasize the use of a convenient, modular, and hybrid quantum pipeline for drug discovery, which will make it more accessible for drug design experts without a quantum computing background. Additionally, referencing established criteria in the drug design domain, our computational results indicate that they also fall within reasonable bounds. Furthermore,} the quantum computing methodologies developed in this study have the potential to extend beyond the presented case studies of Sotorasib and $\beta$-lapachone. The integration of quantum computing into QM/MM simulations offers a versatile platform that can be adapted and scaled to address a wide range of molecular targets and complex biological interactions.

\section{Methods}

\subsection{Quantum Computing for Molecular Systems}
The VQE algorithm uses a parameterized quantum circuit (PQC) $\ket{\psi(\boldsymbol{\theta})}$ to construct a quantum state that approximates the ground state of the system. 
The parameters of the quantum circuit $\boldsymbol{\theta}$ are optimized to its optimal value $\boldsymbol{\theta}^*$ using a classical optimization algorithm, such as gradient descent or Newton's method, to minimize the energy of the quantum state $E(\boldsymbol{\theta})= \braket{\psi(\boldsymbol{\theta})|\hat H|\psi(\boldsymbol{\theta})}$. 
According to the Rayleigh–Ritz variational principle, $E(\boldsymbol{\theta}^{*}) \ge E_{\textrm{ground}}$ and the equity is reached when $\ket{\psi(\boldsymbol{\theta}^*)}$ is the ground state wave function.
Thus, given an expressive PQC, $\ket{\psi(\boldsymbol{\theta}^*)}$ is a good estimation of the ground state wave function.

For molecular systems, the \textit{ab initio} Hamiltonian is written as
\begin{equation}
\label{eq:ham}
 \hat H = \sum_{pq}h_{pq} \hat a^\dagger_p \hat a_q + \frac{1}{2}\sum_{pqrs}h_{pqrs}\hat a^\dagger_p \hat a^\dagger_q \hat a_r \hat a_s \ ,
\end{equation}
where $h_{pq}$ and $h_{pqrs} = [ps|qr]$ are one-electron and two-electron integrals, and $\hat a^\dagger_p, \hat a_p$ are fermionic creation and annihilation operators, respectively.
For chemical systems, the VQE algorithm is composed of several steps. 
The first step is to calculate the integrals in the Hamiltonian under the molecular orbital basis.
Then, the second-quantized fermion Hamiltonian is mapped to a spin Hamiltonian using fermion-qubit mapping,
since quantum computers are built based on the spin model.
In  this work, we employ the parity transformation for saving two qubits
\begin{equation}
    \hat a_j = (\hat c_j \otimes \ket{0} \bra{0}_{j-1} - \hat c^\dagger_j \otimes \ket{1} \bra{1}_{j-1}  ) \otimes \bigotimes_{l={j+1}}^{N-1} \hat X_l 
\end{equation}
Here $\hat c$ is the qubit annihilation operator $\frac{1}{2} (\hat X+i\hat Y)$, and $\hat X$, $\hat Y$ and $\hat Z$ are Pauli operators.
The transformation ensures the preservation of the commutation and anti-commutation properties of fermion operators.
After the fermion-qubit mapping, the Hamiltonian in Eq.~\ref{eq:ham} is transformed to a summation of the products of Pauli operators. More formally, the Hamiltonian can be written as $\hat H = \sum_j^M \alpha_j \hat P_j$ where $\alpha_j$ is the coefficient and $\hat P_j$ is the product of Pauli operators.
$M$ is the total number of terms.
Each $P_j$ can be measured on a quantum computer and subsequently, the overall energy is obtained by taking the weighted summation.

The active-space approximation is employed to reduce computational cost and enhance accuracy.
The approximation adopts the Hartree-Fock state as the baseline state and chooses an ``active space'' that is treated with a high-accuracy computational method.
In classical computation, the high-accuracy method is usually full configuration interaction (FCI) or density matrix renormalization group (DMRG)~\cite{ma2022density}.
In our case, quantum computers are employed to solve the problem with the VQE algorithm.
The active space is usually constructed in the molecular orbital space.
Most commonly, orbitals that have the closest energy with the HOMO and LUMO orbitals will be included in the active space.
Meanwhile, the inner shell orbitals are treated at the mean-field level.
Thus, this approximation is sometimes also called the frozen core approximation.
Denote the set of frozen occupied spin-orbitals by $\Lambda$.
The frozen core provides an effective repulsion potential $V^{\rm{eff}}$  to the remaining electrons

\begin{equation}
\label{eq:effective-pot}
    V^{\rm{eff}}_{pq} = \sum_{m \in \Lambda} \left ( [mm|pq] - [mp|qm] \right).    \ 
\end{equation}

The frozen core also bears the mean-field core energy

\begin{equation}
    E_{\rm{core}} = \sum_{m \in \Lambda} h_{mm} + \frac{1}{2}\sum_{m, n \in \Lambda} \left ( [mm|nn] - [mn|nm] \right ). \
\end{equation}

The \textit{ab initio} Hamiltonian with the active space approximation is rewritten as

\begin{equation}
\label{eq:ham-abinit-as}
    \hat H = \sum_{pq}(h_{pq} + V^{\rm{eff}}_{pq}) \hat a^\dagger_p \hat a_q + \sum_{pqrs}h_{pqrs} \hat a^\dagger_p \hat a^\dagger_q \hat a_r \hat a_s + E_{\rm{core}},  \
\end{equation}

where the indices $p$, $q$, $r$ and $s$ refer to spin-orbitals in the active space.

\subsection{Quantum Computation of Solvation Effect}

The solvation effect is an important topic in classical computational chemistry~\cite{tomasi2005quantum}.
The PCM model is one of the most popular methods to treat the solvation effect~\cite{miertuvs1981electrostatic},
and its combination with VQE has been demonstrated based on a classical emulator~\cite{castaldo2022quantum}.
The PCM model treats the solvent molecules as a continuous homogeneous medium with relative permittivity $\varepsilon_s>1$.
The solvent continuum is polarized by the solute molecule, and in turn, modifies the charge distribution of the solute molecule.
More specifically, the molecule is considered to reside in a van der Waals molecular cavity defined as a union of spheres centered at the atoms

\begin{equation}
    \Omega = \bigcup_{j=1}\Omega_j \ .
\end{equation}

The relative permittivity $\varepsilon(\vec r)=1$ for $\vec r \in \Omega$ and $\varepsilon(\vec r)=\varepsilon_s$ for $\vec r \notin \Omega$.
The additional energy contribution of the electrostatic interaction is
\begin{equation}
    E_s = \frac{1}{2}\int_{\mathbb{R}^3} \rho(\vec r) V_r(\vec r) d \vec r \,
\end{equation}
where $\rho$ is the charge distribution of the solute molecule and $V_r$ is the reaction-field potential by the dielectric continuum.
The reaction field $V_r$ is modeled by a single layer of charges $\sigma(\vec s)$ on the cavity surface $\Gamma = \partial \Omega$
\begin{equation}
    V_r(\vec r) = \int_\Gamma \frac{\sigma(\vec s)}{|\vec r - \vec s|} d\vec s
\end{equation}
In this work, we use quantum computers to model the solute molecule and use classical computers to calculate the solvent potential $V_r$ that is added to the Hamiltonian of the solute molecule.

In our work, we employ one of the variants of PCM, namely the conductor-like screening model (COSMO)~\cite{klamt1993cosmo}, to model the solvation effect on real quantum devices.
COSMO has become very popular due to its ease of implementation, numerical stability, and insensitivity to outlying charge errors.
The COSMO method treats the solvent continuum as a conductor, which simplifies the calculation of $V_r$, and scales $E_s$ by a constant factor $f(\varepsilon_s)$ to take into account the non-conductor nature of the solvents.
In the large $\varepsilon_s$ limit $f(\varepsilon_s)$ should converge to 1.
Based on the conductor model, the surface charge $\sigma(\vec s)$ is obtained by solving the integro-differential equation numerically

\begin{equation}
\label{eq:sol-boundary}
\begin{aligned}
    -\nabla^2 V_r(\vec r) & = 0  \quad &\textrm{for} \ \vec r \in \Omega \ , \\
    V_r(\vec s) & = - \Phi(\vec s) \quad &\textrm{for} \ \vec s \in \Gamma \ . 
\end{aligned}
\end{equation}

Here $\Phi(\vec r)=\int_{\mathbb{R}^3} \frac{\rho(\vec s)}{|\vec r - \vec s|} d\vec s$ is the potential generated by $\rho$ \textit{in vacuo}.
The domain decomposition algorithm is one of the most popular methods to solve Eq.~\ref{eq:sol-boundary}, which offers both high accuracy and high efficiency~\cite{cances2013domain, lipparini2013fast}.
Thus the method is dubbed as ddCOSMO.

The input for solving Eq.~\ref{eq:effective-pot} is the solute charge distribution $\rho$. 
$\rho$ can be computed from the one-body reduced density matrix of the solute molecule.
Similar to the case of computing molecular forces, one-body reduced density matrix 
can be measured on a quantum computer after the main VQE iteration.
After $\sigma(\vec s)$ is determined, the generated potential $V_r$ is added to the Hamiltonian of the solute molecule and effectively modifies $h_{pq}$.
Then, VQE is performed based on the updated Hamiltonian after active-space reduction and yields the updated one-body reduced density matrix.
In quantum computer simulation, we observed that the effect of the iteration is smaller than the measurement uncertainty. Therefore in our quantum computations, we forego iteration and perform only a single calculation.

\subsection{Quantum Computation of Molecular Forces}
Most straightforwardly, the molecular forces can be calculated with numerical finite-difference over the nuclear coordinates. 
Analytical computation, if available, is preferred over such an approach, since analytical computation is both more efficient and accurate.
In our approach, the HF molecular orbital coefficients $\mathbf{C}$ are determined before the VQE calculation. 
As a result, the energy is not stationary to the variation of orbital coefficients $\pdv{E}{\mathbf{C}} \neq 0$ and this term will contribute to the force expression~\cite{helgaker1988analytical}.

In the general form, the force expression can be obtained by chain-rule differentiation as~\cite{taylor1984analytical}
\begin{equation}
\label{eq:force}
\begin{aligned}
    \dv{E}{x}  & = -f_\textrm{nuc} - f_{\textrm{elec}} + \sum_{\mu\nu} (\mu'|\nu) R_{\mu\nu} + 2\sum_{\mu\nu} (\mu'|h|\nu) \left (D^I_{\mu\nu} + D^A_{\mu\nu} \right ) \\
    & \quad + 4\sum_{\mu\nu\lambda\sigma} [\mu'\nu|\lambda\sigma] \times \Big( 2D^I_{\mu\nu} D^I_{\lambda\sigma} - \frac{1}{2} D^I_{\mu\lambda} D^I_{\nu\sigma} - \frac{1}{2} D^I_{\mu\sigma} D^I_{\nu\lambda} \\
& \quad + 2D^I_{\mu\nu} D^A_{\lambda\sigma} - \frac{1}{2} D^I_{\mu\lambda} D^A_{\nu\sigma} - \frac{1}{2} D^I_{\mu\sigma} D^A_{\nu\lambda} \\
& \quad + 2D^A_{\mu\nu} D^I_{\lambda\sigma} - \frac{1}{2} D^A_{\mu\lambda} D^I_{\nu\sigma} - \frac{1}{2} D^A_{\mu\sigma} D^I_{\nu\lambda}  + P^A_{\mu\nu\lambda\sigma}    \Big ) \ .
\end{aligned}
\end{equation}
Here, we have switched from molecular orbital basis to atomic orbital basis, and we use $\mu, \nu, \lambda, \sigma$ instead of $p, q, r, s$ as orbital indices to indicate the different basis.
In Eq.~\ref{eq:force}, $ -f_\textrm{nuc}$ and $(- f_{\textrm{elec}} )$ are the nuclear and electronic Hellmann-Feynman force by $\braket{\psi|\pdv{\hat H}{x}|\psi}$.
Since electron repulsion is invariant to $x$, $\pdv{\hat H}{x}$ is a single-body operator and $\braket{\psi|\pdv{\hat H}{x}|\psi}$ can be computed from one-body reduced density matrix.
$\sum_{\mu\nu} (\mu'|\nu) R_{\mu\nu}$ represents the ``density force'' contribution~\cite{pulay1977direct} which stems from the variation over the orbital coefficients $\mathbf{C}$.
$R_{\mu\nu}$ are the matrix elements of $\mathbf{R}=\mathbf{C}\bm{\epsilon}\mathbf{C}^\dagger$ where $\bm{\epsilon}$ are HF molecular orbital energies.
The remainder of Eq.~\ref{eq:force} represents the ``integral force'' contribution~\cite{pulay1977direct} which stems from the variation over the basis sets.
The primed atomic orbital index in the integrals denotes the derivative of the primed atomic orbital with respect to $x$.
$\mathbf{D}^I$ and $\mathbf{D}^A$ are the one-body reduced density matrices for the inactive and active space respectively
and $\mathbf{P}^A$ is the two-body reduced density matrix for the active space.
Thus, in order to compute the molecular forces on quantum computers, the key is to measure the one- and two-body reduced density matrices of the active space.

We note that it is possible to rewrite Eq.~\ref{eq:force} as the expectation of a ``force operator'' that is formally similar to the \textit{ab initio} Hamiltonian Eq.~\ref{eq:ham}~\cite{o2022efficient}.
As a result, measurement grouping methods developed for energy measurement can be employed directly to reduce the measurement cost~\cite{huggins2021efficient}.
In this study, we do not consider this optimization for ease of implementation.

\subsection{Quantum Computation Details}
We employ the hardware efficient $R_y$ ansatz as the parameterized quantum circuit 
for both the covalent bond simulation and the prodrug activation optimization.
The $R_y$ ansatz is suitable for the simulation of chemical systems since it enforces real amplitudes~\cite{gao2021computational, gao2021applications,mihalikova2022cost, choy2023molecular}.
Compared with the unitary coupled-cluster family of ansatz~\cite{anand2022quantum}, hardware-efficient ansatz requires shorter circuit,
which ensures that the effect of quantum gate noise does not significantly deteriorate our result.
The $R_y$ ansatz is composed of interleaved layers of single-qubit $R_y$ rotation gates and two-qubit CNOT gates
\begin{equation}
    \ket{\psi(\boldsymbol{\theta})}_{R_y} :=  \prod_{l=k}^1 \left [ L_{R_y}^{(l)}(\boldsymbol{\theta}) L_{\rm{CNOT}}^{(l)} \right ] L_{R_y}^{(0)}(\boldsymbol{\theta}) \ket{\phi}, 
\end{equation}
where $k$ is the total number of layers. 
In this study, we employ $k=1$ to reduce the negative impact of the quantum gate noise.
The layers are defined as
\begin{equation}
\begin{aligned}
L_{\rm{CNOT}}^{(l)} & = \prod_{j=N-1}^{1} \textrm{CNOT}[j, j+1], \ \\
 L_{R_y}^{(l)}(\boldsymbol{\theta}) & =  \prod_{j=N}^{1} R_y[{j}](\theta_{lj}). \
\end{aligned}
\end{equation}
Here, $\textrm{CNOT}[j, j+1]$ represents CNOT gate acting on the $j$th and $(j+1)$th qubit,
and $R_y[{j}]$ is $R_y$ rotation gate acting on the $j$th qubit.
$N$ is the total number of qubits.
In our superconducting platform, the $R_y$ gates are compiled into native $R_z$ gates.

The classical emulation of quantum computers is performed using the \textsc{TenCirChem}~\cite{li2023tencirchem} and \textsc{TensorCircuit}~\cite{zhang2023tensorcircuit} package. 
The circuit parameters are pre-optimized on a classical simulator employing the L-BFGS-B optimizer in \textsc{SciPy}~\cite{scipy} and the parameter-shift rule for gradients~\cite{mitarai2018quantum, schuld2019evaluating}.
Due to its efficient architecture, in \textsc{TenCirChem} it takes only a few lines of code to transfer the calculation workflow from classical emulators to real quantum devices.
The solvation energy and molecular forces are calculated classically, after obtaining reduced density matrices on quantum computers, via \textsc{PySCF}~\cite{sun2018pyscf}.

\subsection{Classical Computation Details}

\subsubsection{Methods for obtaining the optimal geometry configuration}

In our C-C bond cleavage based prodrug activation strategy, we should first obtain the optimal geometric configurations of the corresponding molecules to compute the Gibbs free energy profiles. DFT calculations were performed using Gaussian 16. Specifically, we employed the B3LYP functional within DFT, chose the 6-31+G(d) basis set for the molecular orbitals and used Solvation Model Based on Density (SMD) as the solvation model. Throughout the optimization process, we maintained constant connectivity between atoms and applied Grimme's D3 dispersion correction.

\subsubsection{System preparation for covalent bond simulation}

Our simulation started with the intricate system preparation using Amber
software suite \cite{case_ambertools_2023}, especially packages including pdb4amber, antechamber, parmchk2, tleap,
etc, defining the fundamental molecular and environmental parameters. This
initial setup was crucial in modeling the drug-target complex accurately, since
our simulation involves the modified non-standard Amino Acid, in which the Sotorasib
molecule had been glued to the mutated cystein on KRAS. The general process
includes the preprocessing of the protein structure, its split into
different parts, its format conversion, force field parameters generation,
and finally collecting the parts into a complete system ready for the simulation.

In our simulations, QM region was carefully chosen to
include the critical reactive atoms of the KRAS(G12C) mutation. (See Figure \ref{fig:6oim_atoms})  Five atoms that
are key to the stability of the covalent bond (SG on the cysteine side, and C18,
C17, O16 and C15 on the Sotorasib side), have been included in the QM region. 
A covalent bond is formed between the C18 atom on Sotorasib and the SG atom on Cysteine.
Two other atoms, C17, which is covalent bonded to C18, and C15, which is in turn 
covalently bonded to C17, are also included. Another atom, O16, that is sterically 
positioned close to SG and might consequentially affect its atom position and 
bonding, is also included. This meticulous selection allowed for a detailed analysis of the electronic and
structural changes occurring upon drug binding. We also took some inspiration 
from \cite{grigorenko_multiscale_2023} on how to set up the system for covalent bond simulation.

We employed \textsc{OpenMM}~\cite{eastman_openmm_2015} for conducting the molecular dynamics aspect of our study,
while \textsc{PySCF} provided the quantum mechanical calculations essential for
simulating the covalent interactions with high precision. This combination also
guarantees a smooth transition to the later quantum computer implementation,
since our quantum simulation and real machine adaption will be based on \textsc{TenCirChem} and \textsc{PySCF}.

A crucial aspect of our simulation was the calibration of parameters such as
temperature and pressure to replicate physiological conditions accurately, and
considerable care had been taken to formulate a customized Langevin integrator,
to cater to the special energy communications between the QM region and the QC
region of the system. This calibration, along with the integration of a custom
force field, enabled us to capture the nuanced quantum mechanical energies and
forces at play during the formation of the covalent bond between Sotorasib and the
KRAS(G12C) mutation.

\bibliography{main}

\section*{Acknowledgements}
The authors thank Xiaxiaoman Studio for drawing the illustrations. X. Z. thanks grants from the National Natural Science Foundation of China (grant 82322062) and Jiangsu Provincial Funds for Distinguished Young Scientists (grant BK20211527).

\section*{Author contributions statement}
Z. Y., X. Z. and S. Z. conceived and supervised the project. 
W. L., Z. Y., X. L., D. M., S. Y., Z. Z., C. Z., K. B. and M. D. wrote the code and performed the experiments.
W. L., Z. Y., X. L., D. M., S. Y. and Y. C. analyzed the results.
W. L., Z. Y., X. L., D. M., S. Y., J. Y., S. Z. wrote the manuscript.
All authors reviewed the manuscript. 

\section*{Data Availability}
The dataset and results that support the findings of this study are available on GitHub (https://github.com/AceMapAI/qc-drug-design). Source data are provided in this paper.

\section*{Code Availability}
The code that supports the findings of this study is available at GitHub 
(https://github.com/AceMapAI/qc-drug-design).

\section*{Additional information}

The authors declare no competing interest.

\newcolumntype{s}{>{\hsize=.5\hsize}X}
\begin{table}[H]
\caption{Comparison of the energy barrier $\Delta G^{\ddag}$  and Gibbs free energy change $\Delta G$, measured in kcal/mol, for the C-C bond cleavage reaction using classical and quantum computational methods. HF and CASCI are calculated on classical computers, while VQE energies are obtained on a superconducting quantum computer. Both CASCI and VQE calculations are based on a (2e, 2o) active space. For the VQE energy, the mean and standard deviation from 4 independent experiments is reported. The DFT energies are obtained from previous work using M06-2X functional.}
\label{tab:delta_energy}
\centering
\begin{tabularx}{\textwidth}{XXXXX}
\toprule
 & HF & CASCI & VQE & DFT \\
\hline
$\Delta G^{\ddag}$ with solvent & 13.1 & 11.8   & $7\pm5$ & 8.3 \\
$\Delta G$ with solvent         & -41.1 & -49.5 & $-51\pm11$ & -10.8 \\
$\Delta G^{\ddag}$ w/o solvent  & 2.6   & 1.4 & $-4 \pm 7$ & --- \\
$\Delta G$ w/o solvent          & -59.3   & -68.0 & $-65\pm9$ & --- \\
\bottomrule
\end{tabularx}
\end{table}

\begin{table}[H]
\setlength{\tabcolsep}{8pt}
\caption{Gibbs free energy for the molecules studies in this work by classical and quantum computational methods. HF and CASCI are calculated on classical computers, while VQE energies are obtained on a superconducting quantum computer. Both CASCI and VQE calculations are based on a (2e, 2o) active space. For the VQE energy, energy data from 4 independent experiments are reported.}
\label{tab:energy_data}
\centering
\begin{tabular}{lcccc}
\toprule
Molecule & Solvent & HF & CASCI & VQE  \\
\hline
\textbf{4}  &  with solvent   & -1221.2446 & -1221.2447  & -1221.240, -1221.243, -1221.227, -1221.242  \\
            &  w/o solvent    & -1221.1821 & -1221.1821 & -1221.177, -1221.180, -1221.18, -1221.186    \\
\textbf{5} & with solvent     &  -343.3933 & -343.4038  & -343.395, -343.426, -343.403, -343.407 \\
            &  w/o solvent    & -343.3805 & -343.3912   & -343.390, -343.390, -343.391, -343.390     \\
\textbf{6} & with solvent     &  -801.8703 & -801.8732 & -801.866, -801.878, -801.859, -801.866 \\
            &  w/o solvent    & -801.8576 & -801.8607   & -801.856, -801.839, -801.857, -801.852      \\
\textbf{TS} & with solvent    & -1221.2238 & -1221.2259 & -1221.224, -1221.232, -1221.223, -1221.230  \\
            &  w/o solvent    & -1221.178 & -1221.180 & -1221.181, -1221.187, -1221.179, -1221.207     \\
\ce{H2O}     & with solvent   & -76.0465 & -76.0466 & -76.043, -76.059, -76.036, -76.041\\
            &  w/o solvent    & -76.0385 & -76.0386 & -76.029, -76.035, -76.059, -76.053     \\
\bottomrule
\end{tabular}
\end{table}

\begin{table}[H]
\caption{Comparison of computational wall times for classical computing (CASCI) and quantum computing (VQE) on solving the active space of molecule \textbf{4}, \textbf{5}, \textbf{6}, and \textbf{TS}.}
\label{tab:time_cmp}
\centering
\begingroup
\setlength\tabcolsep{16pt}
\begin{tabular}{lccc}
\toprule
& \multicolumn{2}{c}{Computational Wall Time (s)} \\
  Molecule    & CASCI & VQE \\
\midrule
\textbf{4} &  358 & 424 \\
\midrule
\textbf{5} & 3 & 63 \\
\midrule
\textbf{6} & 97 & 161 \\
\midrule
\textbf{TS} & 360 & 424 \\
\bottomrule
\end{tabular}
\endgroup
\end{table}

\begin{table}[H]
\caption{Comparison of simulation times for three different experiments: a classic QM/MM simulation, a noiseless quantum emulation, and a quantum computer simulation. The total time is for 1600 steps, with the average time per step calculated accordingly.}
\label{tab:compare}
\centering
\begin{tabular}{lrr}
\toprule
   Method &  Total Time (minutes) &  Avg Time per Step (seconds) \\
\midrule
  Classic &                 124.0 &                      4.65 \\
Noiseless &                 126.0 &                      4.725 \\
  Quantum &                3820.0 &                      143.25 \\
\bottomrule
\end{tabular}
\end{table}

\end{document}